\documentclass[runningheads]{llncs}
\usepackage[svgnames]{xcolor}
\usepackage{graphicx}
\usepackage{todonotes}
\usepackage{cite}
\usepackage{subcaption}
\usepackage{algpseudocode}
\usepackage{algorithm}

\usepackage{siunitx}

\usepackage{amsmath}
\usepackage{amsfonts}
\usepackage{amssymb}

\usepackage{hyperref}
\hypersetup{
  colorlinks=true,
  linkcolor=DarkBlue,
  citecolor=DarkGreen,
  urlcolor=DarkMagenta,
}

\usepackage{tikz}
\usetikzlibrary{decorations.pathreplacing,shapes,arrows,positioning}
\tikzset{
	vertex/.style = {
		circle,
		fill            = black,
		outer sep = 2pt,
		inner sep = 1pt,
	}
}
\usetikzlibrary{spy}
\usetikzlibrary{matrix}

\DeclareMathOperator*{\argmin}{arg\,min}
\usepackage{xspace}
\newcommand*\MMLEM{M-ML-EM\xspace}
\begin{document}
\title{Spatiotemporal PET reconstruction using ML-EM with learned diffeomorphic deformation}
\titlerunning{Learned spatiotemporal PET reconstruction}
%
\author{Ozan \"Oktem%
\inst{1}%
\and
Camille Pouchol%
\inst{1}%
\and
Olivier Verdier%
\inst{1,2}%
}
\authorrunning{O. \"Oktem et al.}
%
\institute{Department of Mathematics, KTH Royal Institute of Technology, 100 44 Stockholm, Sweden \and
Department of Computing, Mathematics and Physics, Western Norway University of Applied Sciences, Bergen, Norway
}

%
\maketitle              
\begin{abstract}
  Patient movement in emission tomography deteriorates reconstruction quality because of motion blur.
  {Gating} the data improves the situation somewhat: each gate contains a movement phase which is approximately stationary.
  A standard method is to use only the data from a few gates, with little movement between them.
  However, the corresponding loss of data entails an increase of noise.
  Motion correction algorithms have been implemented to take into account all the gated data, but they do not scale well, especially not in 3D.
  We propose a novel motion correction algorithm which addresses the scalability issue.
  Our approach is to combine an enhanced ML-EM algorithm with deep learning based movement registration.
  The training is unsupervised, and with artificial data.
  We expect this approach to scale very well to higher resolutions and to 3D,
  as the overall cost of our algorithm is only marginally greater than that of a standard ML-EM algorithm.
  We show that we can significantly decrease the noise corresponding to a limited number of gates.
\end{abstract}

\section{Introduction}

Positron emission tomography (PET) is a molecular imaging technology where a radioactive tracer is administered to a patient. 
The tracer is an x-ray source that emits pairs of photons travelling into opposite directions, and the PET scanner is an arrangement of detectors for detecting such photon pairs (coincidence events).
The goal is then to recover the spatial distribution of the tracer (activity map) from these coincidence events.

Acquiring a sufficient amount of coincidence events takes time, typically twenty to forty minutes depending on the detector efficiency and the size of the region being imaged.
Organs, such as the heart and lungs, move during the PET data acquisition, so the activity map one seeks to recover in PET imaging is a spatiotemporal quantity. 
Failure to account for the temporal variability during reconstruction results in a deteriorated PET image.

\subsection{Survey of existing works}
Most approaches that consider motion in PET image reconstruction assume access to gated PET data. 
Here, PET data is subdivided into subsets where the coincidence data is from the activity map in a specific temporal state. 
For cardiac and respiratory motion, gated data would correspond to decomposing the entire dataset into
parts that represent different breathing and/or cardiac phases. 
Hence, the activity associated to each gate can be assumed to be stationary, but data in the gates also suffer from a relatively low signal-to-noise ratio since they only contain a small portion of the coincidence events.

%
%


A straightforward approach based on gated data is to recover each temporal state of the activity independently of each other (frame-by-frame reconstruction). 
This does not account for the temporal dynamics of the activity, instead it reduces the spatiotemporal reconstruction problem into a sequence of independent stationary reconstruction problems, which in PET is done by ML-EM~\cite{ShVa82} (or a variant thereof, like OSEM~\cite{Hudson1994}).  
Spatiotemporal reconstruction refers to methods that instead take the temporal dynamics into account.
Several approaches have been proposed where most rely on estimating a motion model prior to reconstruction, see \cite{DaJiSc08,RaTaZa13,ReVe14,GiJiDaSc15} for survey. 




In this paper, the proposed method falls into the family of algorithms that, contrarily to those based on a priori built motion models, jointly estimate the image and motion, directly from the full set of measured data. 
An objective function is optimised with respect to
two arguments: image and motion. Hence, only one image with the full statistic is reconstructed. 
Given the close relationship between the image reconstruction and motion estimation steps, a simultaneous method of estimating 
the two is better able to reduce motion blur and compensate for poor signal-to-noise ratios and to 
improve the accuracy of the estimated motion \cite{GiMaBoal02,GrYaKiJi06}.


In the latter works, one performs a two-step minimisation of a joint energy functional term 
(which includes both image likelihood and motion-matching terms). 
The method chosen by Jacobson and Fessler\cite{JaFe03,JaFe06}, referred to as \textit{joint estimation with deformation modelling}, is based on maximising the likelihood for a parametric Poisson model for gated PET measurements. Motion (from gate to gate) is defined by a set of deformation parameters. 
A similar motion-aware likelihood function was used by Blume and colleagues \cite{BlMaKeNaRa10}, although using a distinct optimisation scheme and depicting more convincing results. In this context one may also consult \cite{ZhGhBr05}, which compares three approaches for joint reconstruction of image and motion.

An alternative is to consider motion models derived from deformations modelled by diffeomorphisms, as obtained from example through the LDDMM framework \cite{Yo10}.
Here, one can calculate regularising functionals that incorporate such deformations. Finally, \cite{FaShElFa14} provides an overview of variational 
shape models as applied to the registration and segmentation problems. These could also be coupled with variational 
regularisation methods for image reconstruction. 


The main drawback of all these methods, however, is the relatively high computational 
costs involved in the joint reconstruction approach. 

\subsection{Proposed method}
In this paper, we develop a joint reconstruction method based on the minimisation of a suitable functional.
The main novelty of our work is the scalability of the resulting algorithm, as its complexity is of the order of the usual ML-EM algorithm.
Images are indeed estimated using a generalised ML-EM algorithm.
Motion estimation, with deformations modelled by diffeomorphisms, is based on the unsupervised deep learning framework \texttt{voxelmorph}~\cite{Dalca2018}.
That is, we make use of a pre-trained neural network which performs direct image registration, i.e., the network finds a diffeomorphism which, given two images, deforms the first one into the second. 

Interestingly, one single outer iteration of our algorithm is close to the recently proposed approach~\cite{Li2019}. Thus, it generalises the previous work and shows that it can be interpreted in the framework of an optimisation problem.

The results of the proposed method are tested on the Derenzo phantom, and shown to recover a significant part of the information lost when one uses gate-by-gate reconstruction.

\section{Methods}
\label{Methods}
\subsection{Mathematical background}

\subsubsection{ML-EM algorithm~\cite{ShVa82}.}
Let us consider the statistical model
\[g \sim \textrm{Poisson}(Af),\]
where $f$ is the unknown image, and $g$ is the acquired data---a vector of $\mathbb{R}^d$;
this models the physics of \emph{stationary} PET with forward operator $A$.

The ML-EM algorithm solves the corresponding \emph{maximum likelihood problem}, which amounts to minimising the divergence $d_{KL}(g|| Af)$, defined for two non-negative vectors $u$, $v$ in $\mathbb{R}^d$ by
\[d_{KL}(u||v) := \sum_{j=1}^d \left(v_j - u_j - u_j \log\Big(\frac{u_j}{v_j}\Big)\right).\] 
The ML-EM algorithm is an iterative solver with update
\begin{equation}
\label{mlem}
f^{(n+1)} := \frac{f^{(n)}}{A^T 1}\, A^T\bigg(\frac{g}{A f^{(n)}}\bigg),
\end{equation}
starting from an initial guess $f^{(0)}$, usually $f^{(0)} = 1$.  

\subsubsection{Diffeomorphisms acting on images.} 
Viewing images as elements of $X:= L^2(\Omega)$, {i.e.}, square-integrable functions on a compact $\Omega \subset \mathbb{R}^p$ with $p=2$ or $p=3$, we model motion as an appropriate \emph{group action} of diffeomorphisms onto images.
In this paper,
given a diffeomorphism $\psi \colon\Omega\to\Omega$,
we will use the specific definition $\mathcal{W}_\psi : X \mapsto X$
as
the \emph{intensity-preserving} action \[\mathcal{W}_\psi f(x) := f(\psi^{-1}(x)).\]
Note that our approach is, however, general, and we could have used the \emph{mass-preserving} action instead, namely
\begin{equation}
  \label{eq:densityrep}
  \widetilde{\mathcal{W}}_\psi f(x) := |D\psi^{-1}(x)| f(\psi^{-1}(x))
  .
\end{equation}

We will parameterise diffeomorphisms by exponentials of (stationary) vector fields, i.e.,
\(\psi = \exp(v)\), 
where the exponential $\exp(v)$ of a vector field $v$ is defined as $\psi(1, \cdot)$, where $\psi(t, \cdot)$ solves the differential equation \(\frac{\partial \psi}{\partial t}(t,\cdot) = v(\psi(t, \cdot))\), with initial condition $\psi(0, \cdot) = \operatorname{Id}$. 

\subsubsection{Image registration.} 

The (direct) image registration problem consists in deforming a \emph{template} $f_1$ into a \emph{target} $f_2$, {i.e.}, finding a diffeomorphism $\psi$ such that $\mathcal{W}_\psi f_1 \approx f_2$.
This is usually done by minimising a functional of the form
\begin{equation}
\label{motion_est_2}
\argmin_{\psi} \; d_2(f_2, \mathcal{W}_{\psi} f_1) + \lambda \mathcal{R}(\psi), 
\end{equation}
where $d_2$ is the $L^2$-distance on $X$, $\mathcal{R}$ is a regularisation term on diffeomorphisms that is discussed in~\autoref{mot_est}, and $\lambda$ is a regularisation parameter. 

\subsection{General approach}

\subsubsection{Modelling.}
We are given \emph{gated} data in $N+1$ different gates, corrupted by Poisson noise. For $g_i$ denoting the data, $f_i$ the images in each gate and $A$ the forward operator, we thus assume 
\[g_i \sim \textrm{Poisson}(Af_i), \; i = 0, \ldots, N.\]

We also assume that for $i = 1,\ldots,N$, two consecutive images $f_{i-1}$ and $f_{i}$ are related by
the statistical model
\[ f_i = \mathcal{W}_{\psi_i} f_{i-1} + e_i,\] 
where $\psi_i\colon\Omega\to\Omega$ is the exponential of a vector field following a given probability law (see~\eqref{motion_est_3})
and $e_i$ is a $X$-valued random variable.



\subsubsection{Variational problem.}
We now define the variational problem associated to the inverse problem of finding both the images $f_i$ and diffeomorphisms $\psi_i$ from the data $g_i$. It reads
  \begin{equation}
    \label{eq:varprob}
    \argmin_{(f_i), (\psi_i)} \; J(f_0, \ldots, f_N, \psi_1, \ldots, \psi_N)
    ,
  \end{equation}
 where
 \begin{equation*}
   J(f_i, \psi_i) :=\sum_{i=0}^N d_{KL}(g_i || Af_i) + \sum_{i=1}^N \Big(d_2(f_i, \mathcal{W}_{\psi_i} f_{i-1}) + \lambda \mathcal{R}(\psi_i)\Big).
\end{equation*}

\subsubsection{General algorithm.}

We solve the variational problem \eqref{eq:varprob} by an \emph{intertwined} method,
which consists in alternating between estimating the diffeomorphisms (the \emph{motion estimation step}), and the images $f_i$ (the \emph{reconstruction step}).

The images are first initialised by solving the maximum likelihood problem $\arg\min_{f_i} (d_{KL}(g_i || Af_i))$, associated to $g_i = \textrm{Poisson}(Af_i)$ in each gate. This is done by the algorithm ML-EM~\eqref{mlem}, yielding estimates $f_i^0$, $i = 0, \ldots, N$.

For a given estimate of images $f_i^k$, the motion estimation part consists in solving  
\[ \argmin_{(\psi_i)} \;  \sum_{i=1}^N \left(d_2(f_i^k, \mathcal{W}_{\psi_i} f_{i-1}^k)+ \lambda \mathcal{R}(\psi_i)\right), \]
which in turn can be decomposed into $N$ problems of the form 
\begin{equation}
\label{motion_est}
\argmin_{\psi_i} \;  d_2(f_i^k, \mathcal{W}_{\psi_i} f_{i-1}^k) + \lambda \mathcal{R}(\psi_i), \; i=1, \ldots, N.
\end{equation}
Note that each of these  becomes an \emph{image registration problem}, as we are looking for a diffeomorphism $\psi_i^{k+1}$ matching the template $f_{i-1}^k$ against the target $f_i^k$.

For the reconstruction part, we assume $f_i^k  \approx \mathcal{W}_{\psi_i^{k+1}} f_{i-1}^k$ for $i=1, \ldots, N$ and neglect the $N$ corresponding $d_2$ terms.
The minimisation problem thus becomes
\[\argmin_{(f_i)} \;  \sum_{i=0}^N d_{KL}(g_i || Af_i).\]
We then focus on a particular gate, say the zero'th gate,
and use $f_i^k  \approx \mathcal{W}_{\psi_i^{k+1}} f_i^k$ to obtain
 the optimisation problem with $f_0$ as the only variable:
\begin{equation}
\label{reconst}
\argmin_{f_0}\;  \sum_{i=0}^N d_{KL}\big(g_i || A_i f_0 \big) \big).
\end{equation}
where
\begin{equation}
  \label{eq:defAi}
  A_i := A \mathcal{W}_{\phi_i}
\end{equation}
and we have used the notation $\phi_i := \psi_i \circ \cdots \circ \psi_1$ for $i=1, \ldots,N$.
Solving the above yields a next estimate $f_0^{k+1}$ for $f_0$. All the images $f_i^{k+1}$ are then obtained by $f_i^{k+1} = \mathcal{W}_{\psi_i^{k+1}} f_{i-1}^{k+1}$, $i=1, \ldots, N$. 

It now only remains to explain how the optimisation problems~\eqref{motion_est} and~\eqref{reconst} are solved, which is the topic of the next subsections.

\subsection{Motion estimation}
\label{mot_est}
The motion estimation problem~\eqref{motion_est}, can be rewritten for two generic images $f_1$ and $f_2$ as
\begin{equation}
\label{motion_est_3}
\argmin_{v} \; d_2(f_2, \mathcal{W}_{\exp(v)} f_1) + \lambda \mathcal{R}(v)
,
\end{equation}
where we parameterise the diffeomorphisms by exponentials of stationary vector fields $v$.

To solve this direct image registration problem, we use the \texttt{voxelmorph} unsupervised deep learning approach~\cite{Dalca2018},
where a neural network parameterises a function $(f_1, f_2) \mapsto v$.
That neural network is itself based on the network architecture \texttt{Unet}~\cite{Ronneberger2015}.
 We keep the architecture of \texttt{voxelmorph}, with the same hyperparameters and specific regularisation functional $\mathcal{R}$ given in~\cite{Dalca2018}.
Once trained, the network produces a mapping matching any two images $f_1, f_2$, which we denote
\begin{equation}
  \label{eq:network}
  \gamma(f_1,f_2):= \exp(v(f_1,f_2))
  .
\end{equation}

\textbf{Training.} In~\cite{Dalca2018}, the network is trained on tuples of images $(f_1, f_2)$ coming from brain MRI scans.
We use instead \textit{synthetic} data: tuples of images $(f_1, f_2)$ generated on the fly. 

\begin{figure}[h]
	\centering
	\begin{subfigure}[]{0.32\textwidth}
		\includegraphics[width=\textwidth]{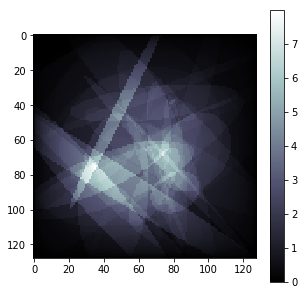}
		\caption{}
	\end{subfigure}
	\begin{subfigure}[]{0.32\textwidth}
		\includegraphics[width=\textwidth]{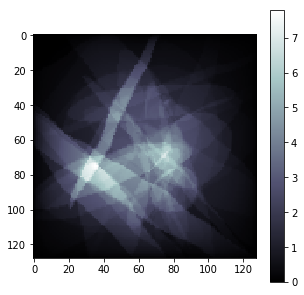}
		\caption{}
	\end{subfigure}
		\begin{subfigure}[]{0.32\textwidth}
		\includegraphics[width=\textwidth]{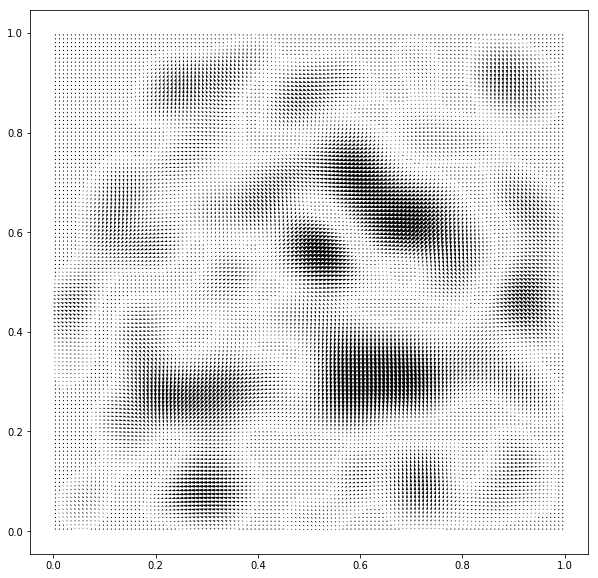}
		\caption{}
	\end{subfigure}
	\\[-2mm]
	\caption{Example of a 2D synthetic tuple of images $f_1$ (a) and $f_2$ (b), related by $f_2 = \mathcal{W}_{\exp(v)} f_1$ for the intensity-preserving action, with $v$ plotted in (c).}
	\label{synthetic}
\end{figure}

We generate training images as follows.
A random image $f_1$ consists of a Poisson random number of ellipsoids~\cite{Adler2017, Adler2018}.
The centre of each ellipsoid has uniform distribution inside the central part of the domain $\Omega$,
the principal axes have exponential distribution,
and the orientation follows a uniform distribution.
We apply a mask vanishing at the boundary to avoid boundary effects.
when diffeomorphisms are applied.
We generate random vector fields $v$ using a Gaussian random field with radial basis function kernel, with appropriate scale and typical size.
The training image $f_2$ is then $f_2 = \mathcal{W}_{\exp(v)} f_1$.
We show in \autoref{synthetic} a sample of images $f_1$, $f_2$ and vector field \(v\) generated as above.


\subsection{Reconstruction}
We now focus on the reconstruction problem~\eqref{reconst}
which we solve
using a reformulation of ML-EM.
Given operators $A_i$, we can simply write ML-EM for the compound operator $A = (A_0,\dotsc,A_N)$ which yields
\begin{equation}
  \label{mlem_diff} f_0^{(n+1)} = \frac{ f_0^{(n)}}{\sum_{i=0}^N  {A_i^T 1}}\sum_{i=0}^N    \, A_i^T \Bigg(\frac{g_i}{A_i  f_0^{(n)}}\Bigg)
  ,
\end{equation}
for an initial guess $f_0^{(0)}$.
We call this algorithm ``\MMLEM'' to avoid the confusion with the vanilla ML-EM algorithm \eqref{mlem}.
We use this algorithm with $A_i$ defined in \eqref{eq:defAi}.
Note that this algorithm has been used in~\cite{Hinkle2012} for the particular case of the intensity-preserving action.
The computation of $A_i^T$ requires the computation of $\mathcal{W}_{\phi_i}^T$.
We achieve this by using the identity $W_{\phi}^T = \widetilde{\mathcal{W}}_{\phi^{-1}}$ valid for any diffeomorphism $\phi$, where $\widetilde{\mathcal{W}}$ denotes the mass-preserving action \eqref{eq:densityrep}.



\subsection{Full algorithm}
We summarise the algorithm with all the necessary details in \autoref{alg:mainalgorithm}.

\begin{algorithm}
  \caption{Full Algorithm}
  \label{alg:mainalgorithm}
  Choose the outer number of iterates $n_{\mathrm{outer}}$, the inner number of iterates $n_{\mathrm{inner}}$ for \MMLEM, and $n_{\mathrm{init}}$, the number of iterates for vanilla ML-EM.
  \begin{algorithmic}
    \For{$i \gets 0,\dotsc,N$}
    \State $f_i \gets \Call{ML-EM}{A, g_i, n_{\mathrm{init}}}$
    \Comment{Iterates of \eqref{mlem}}
    \EndFor
    \For{$k \gets 1,\dotsc,n_{\mathrm{outer}}$}
    \For {$i \gets 1,\dotsc,N$}
    \State $\psi_i \gets \gamma(f_{i-1}, f_{i})$ \Comment{Network registration \eqref{eq:network}}
    \EndFor
    \State $W_0 \gets \mathrm{Id}$
    \For {$i \gets 1,\dotsc,N$}
    \State $W_i \gets W_{\psi_i} W_{i-1}$ 
    \State $A_i \gets A W_i$
    \EndFor
    \State $A_0 \gets A$
    \State $f_0 = \Call{\MMLEM}{\{A_j\}_{j=0,\dotsc,N}, \{g_j\}_{j=0,\dotsc,N}, n_{\mathrm{inner}}}$
    \Comment{Iterates of \eqref{mlem_diff}}
    \For{$i \gets 1,\dotsc,N$}
    \State{$f_i \gets W_i f_0$}
    \EndFor
   
    \EndFor
  \end{algorithmic}
  The outcome is $f_0$.
\end{algorithm}





\subsection{Complexity} 
Evaluating vector fields with the network is negligible when compared to ML-EM or \MMLEM iterations.
Each iteration is itself controlled by the time $\mathsf{t}$ required to compute an expression of the form $A^T(\frac{g}{Af})$.
Since \MMLEM sums these quantities $N$ times, an iteration of it is of the order of $N \times \mathsf{t}$.
Note that evaluating the denominator in \eqref{mlem_diff} (which involves sums of $A^T 1$) does not take more time than evaluating the denominator in ML-EM since $A^T 1$ can be computed off-line.



\section{Results}

\subsection{Derenzo phantom}
\begin{figure}[h]
	\centering
	\begin{subfigure}[]{0.23\textwidth}
		\includegraphics[width=\textwidth]{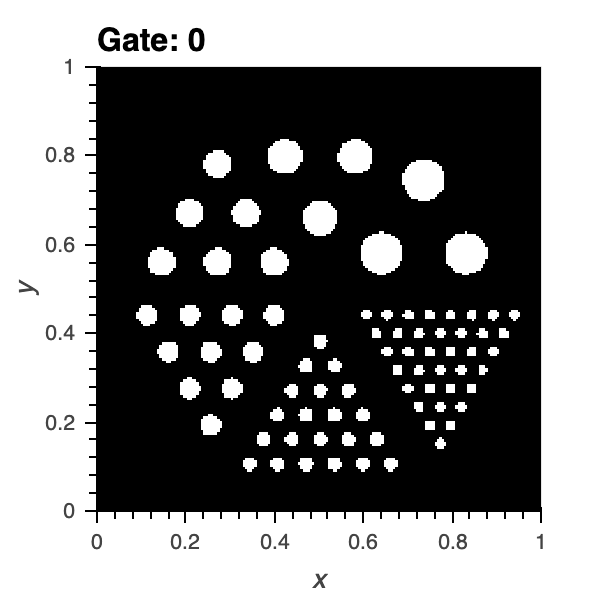}
    \caption{}
    \label{derenzo}
	\end{subfigure}
	\begin{subfigure}[]{0.23\textwidth}
		\includegraphics[width=\textwidth]{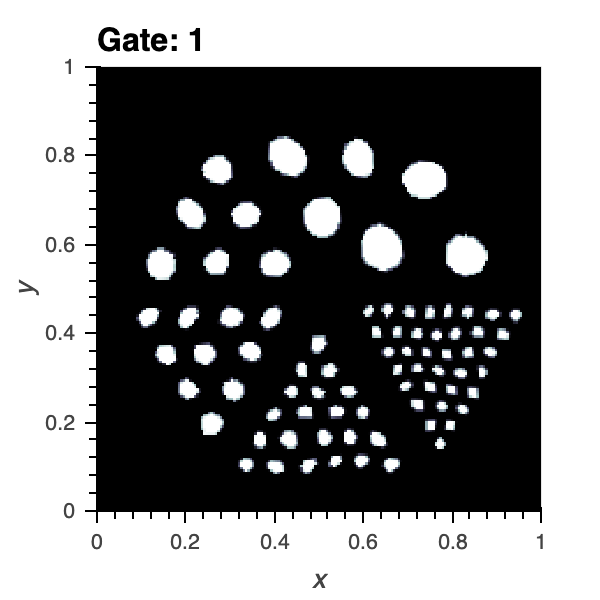}
	\caption{}
	\end{subfigure}
		\begin{subfigure}[]{0.23\textwidth}
		\includegraphics[width=\textwidth]{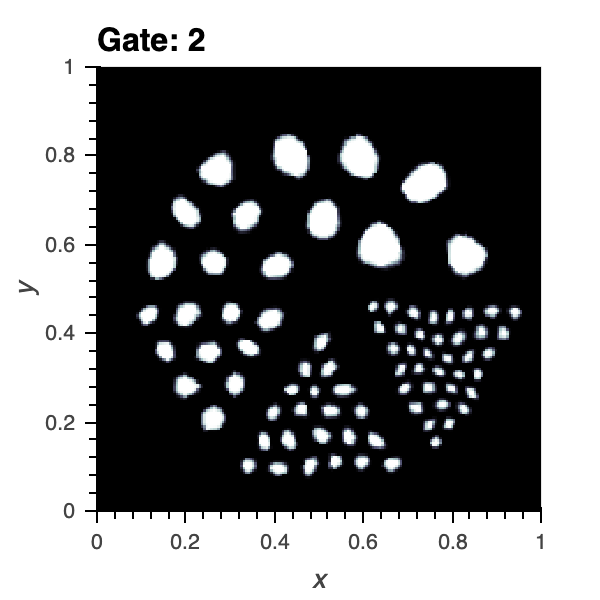}
	\caption{}
	\end{subfigure}
		\begin{subfigure}[]{0.23\textwidth}
		\includegraphics[width=\textwidth]{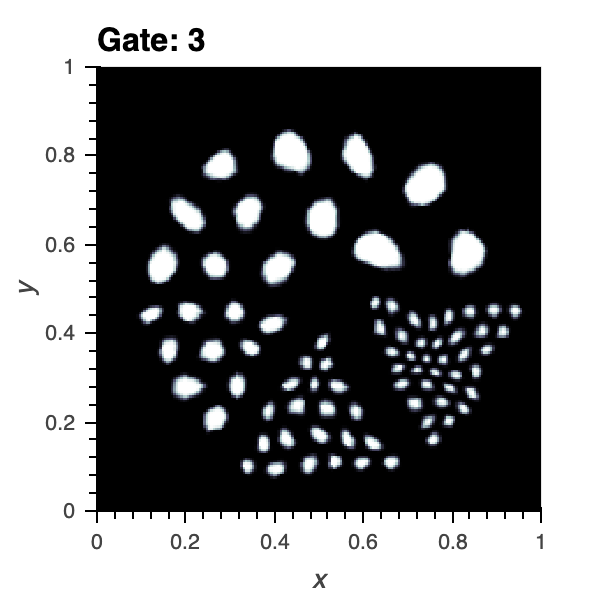}
	\caption{}
	\end{subfigure}
	\\[-2mm]
	\caption{Derenzo phantom in four different gates.}
	\label{phantom}
\end{figure}

We present experiments with the Derenzo phantom, with image size $192 \times192$.
Although this phantom is made of ellipses, we stress that they are very different from the data used to train the network, compare \autoref{synthetic} and \autoref{derenzo}. 

This phantom is then deformed successively with the intensity preserving action by exponentials of vector fields, where each vector field is drawn from the same distribution used to train the network.
For the experiments, we use $N=3$, which amounts to four gates, and we want to recover the image in the initial gate.
The resulting four phantoms are presented in \autoref{phantom}.

The forward operator $A$ is a 2D PET operator with $108$ angles (views) and $250$ tangential positions.
The noisy data is $\mathrm{Poisson}(A(tf))$ for each image $f$, where $t$ is the acquisition time and thus controls the noise level. 

For the phantoms in \autoref{phantom}, we choose $t=60$.
This noise level gives rise to typical optimal numbers of iterates for ML-EM which are of the same order of magnitude as the ones in clinic applications.
Note that all images are multiplied by the same time factor, which amounts to assuming that acquisition time is roughly the same in each gate.

\subsection{Methods without motion correction}

We compare our method with two simple reconstruction methods (simple because without motion correction) for images with gated data:

\begin{itemize}
\item 
  Either one aggregates the whole data and reconstructs from ML-EM, leading to blurry results because of the movement.
  \item
    Or one tries and limit blur by focusing on one gate (say the first) and reconstructing only from that.
    Since there is less data, the result is noisier. 
\end{itemize}

In order to quantitatively compare these strategies, we use ML-EM for the data obtained from taking gate zero only, aggregating gates zero and one, and so on up until aggregating all the four gates.
Finally, we can also estimate the best reconstruction one could hope for, that is, if there were no movement.
This amounts to acquiring the phantom in the $0$th gate four times longer. 

\begin{figure}[h]
	\centering
		\includegraphics[width=0.7 \textwidth]{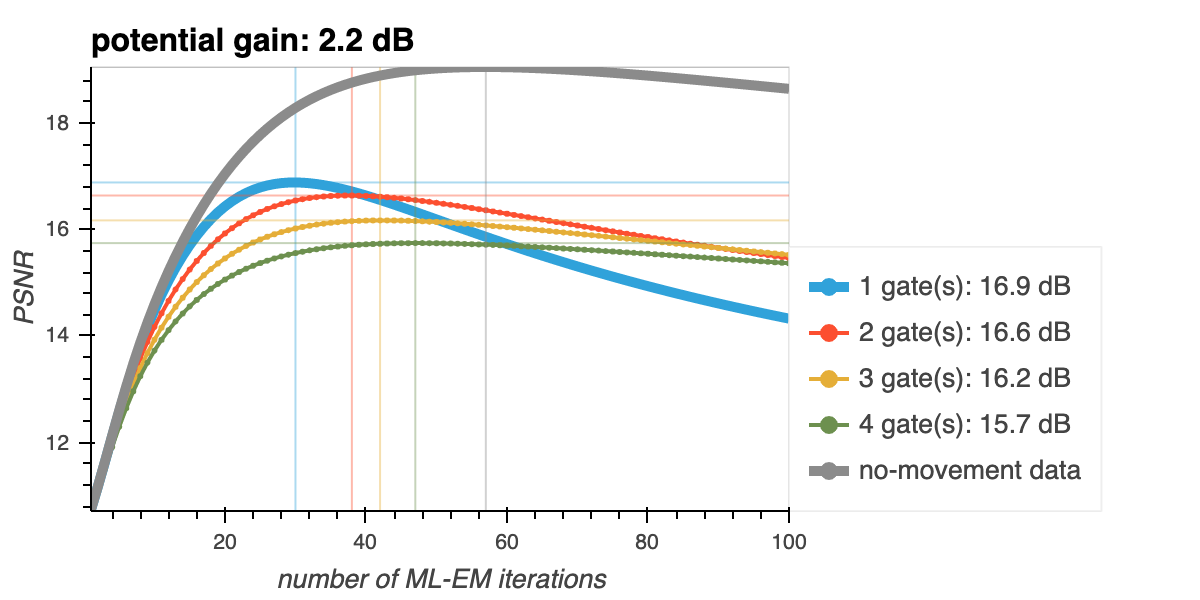}
	\caption{PSNR for different ML-EM strategies without motion correction, and comparison with "no-movement" data, reconstructing from the initial gate acquired four times longer.}
	\label{MLEM_only}
\end{figure}

The results are given in \autoref{MLEM_only}, where the PSNR between the estimated image and the real image in gate zero is computed at each iteration.

The results show that aggregating the gates progressively induces a drop in image quality, as measured by the PSNR.
Compared to gate zero acquired four times longer, the best possible achievable gain is about \SI{2.2}{\decibel}.

\subsection{Proposed method}
We apply \autoref{alg:mainalgorithm} to the data above.
It turns out that a single outer iteration is responsible for most of the improvement, so we focus on that case for presenting experiments.
In other words:
\begin{enumerate}
\item 
we initialise by running some ML-EM iterations in each gate,
\item
we then match the resulting images to estimate the diffeomorphisms,
\item
we finally run some \MMLEM iterations.
\end{enumerate}

\begin{figure}[h]
	\centering
		\includegraphics[width=0.9 \textwidth]{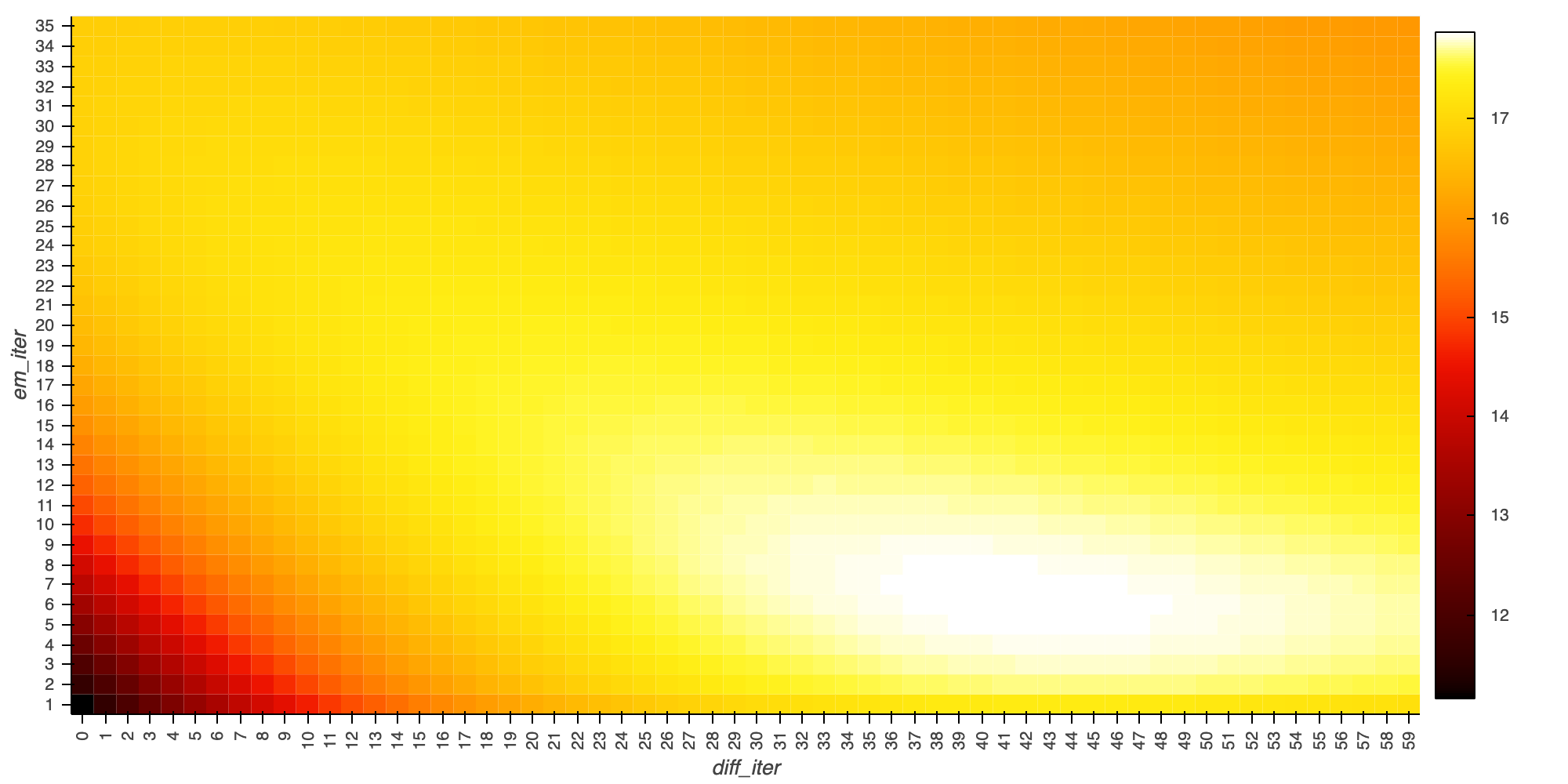}
	\caption{PSNR for various choices of number \texttt{em\_iter} of initial ML-EM iterates  and number \texttt{diff\_iter} of \MMLEM iterates.}
	\label{grid}
\end{figure}

We plot the PSNR between the image reconstructed (in the initial gate zero) and the real image, for a given number \texttt{em\_iter} of ML-EM iteration followed by a given number \texttt{diff\_iter} of \MMLEM iterations.
These results are presented in \autoref{grid}.

We find that the optimal strategy is to iterate only a few times (six iterations in this specific experiment) with ML-EM before estimating the diffeomorphisms through \MMLEM ($42$ iterations in this specific experiment).
Note that this yields a total of $48$ iterations which is higher than the $29$ ML-EM iterations which would be optimal for reconstructing from the gate zero. 

\begin{figure*}
		\centering	
		\begin{subfigure}[t]{0.328\linewidth}
			\centering	
			\begin{tikzpicture}[
			remember picture,
			spy using outlines={%
				circle,
				red,
				magnification=3,
				size=2cm,
				connect spies,
				spy connection path={\draw[thick] (tikzspyonnode) -- (tikzspyinnode);}
			}
			]
			\node {\includegraphics[width=\linewidth, trim={22.5mm 17mm 27mm 6mm}, clip]{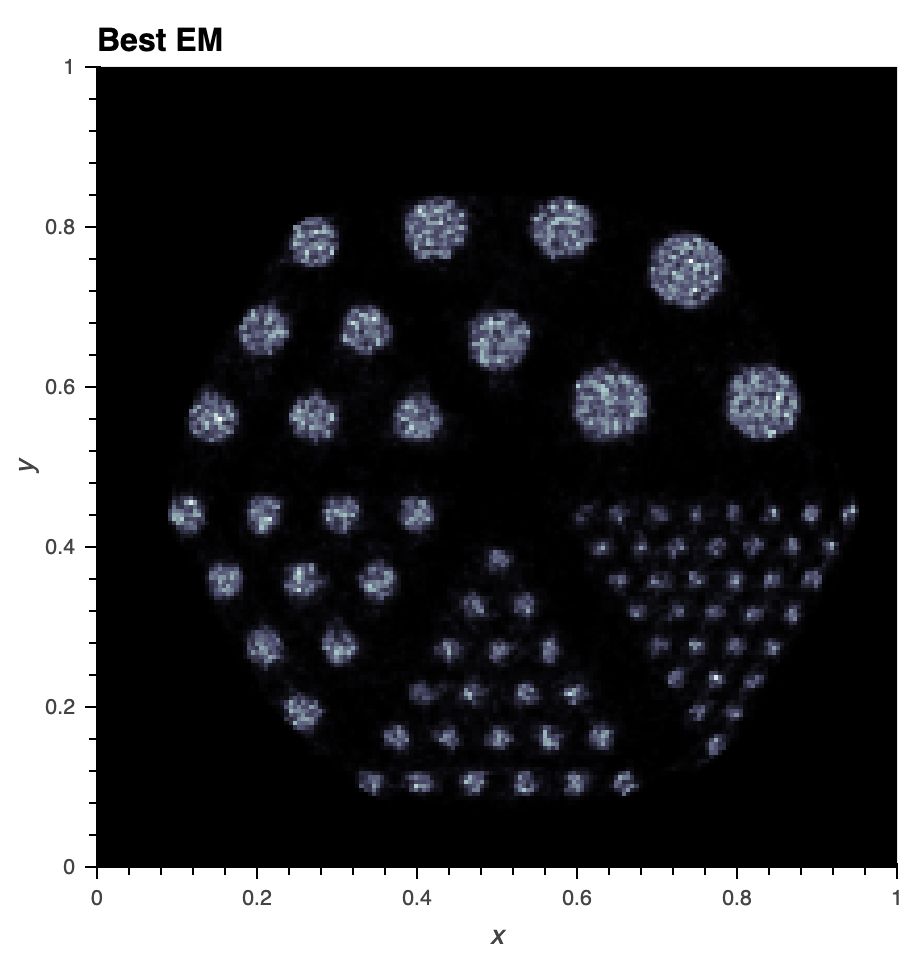}};
			\spy on (0.85,-0.4) in node [left] at (3.2,2.2);
			\end{tikzpicture}
			\caption{Optimal ML-EM reconstruction: 29 iterations (using one gate)}
		\end{subfigure}
		\hspace{1cm}
		\begin{subfigure}[t]{0.328\linewidth}
			\centering		
			\begin{tikzpicture}[
			remember picture,
			spy using outlines={%
				circle,
				red,
				magnification=3,
				size=2cm,
				connect spies,
				spy connection path={\draw[thick] (tikzspyonnode) -- (tikzspyinnode);}
			}
			]
			\node {\includegraphics[width=\linewidth, trim={22.5mm 17mm 27mm 6mm}, clip]{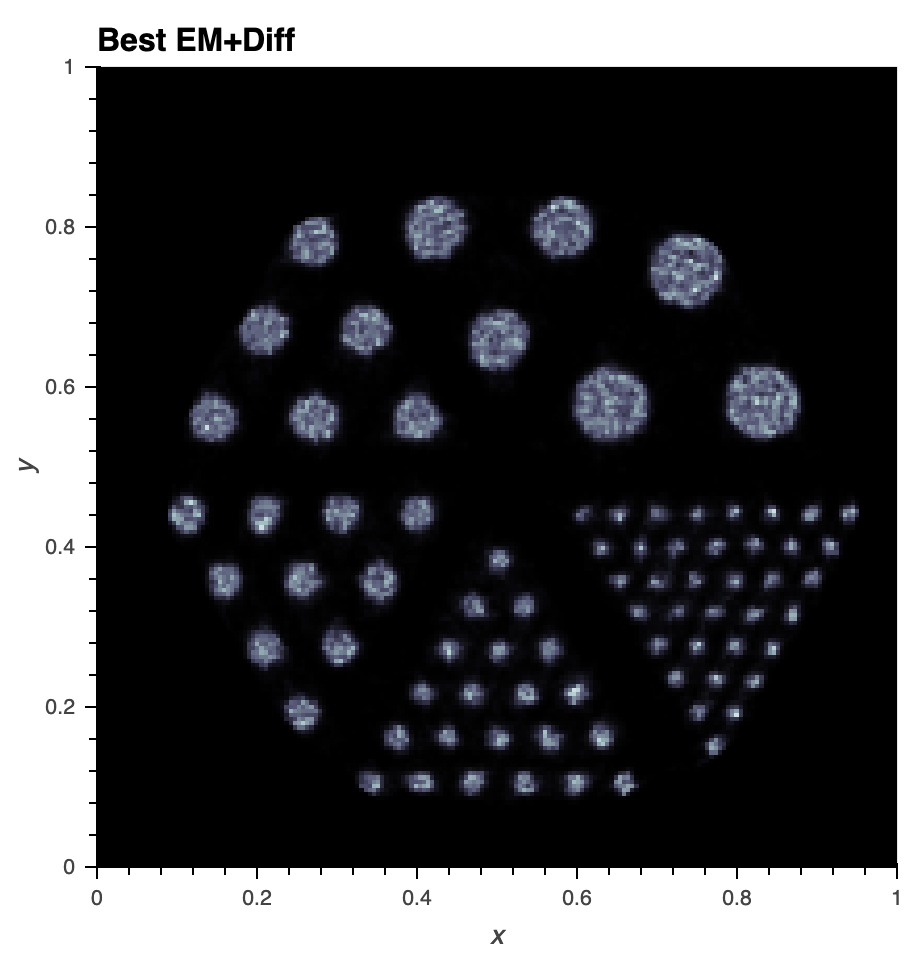}};
			\spy on (0.85,-0.4) in node [left] at (3.2,2.2);
			\end{tikzpicture}
			\caption{Optimal reconstruction with \MMLEM: 6 ML-EM + 42 \MMLEM iterations}
		\end{subfigure}
		\caption{
      Optimal reconstructions of the gate zero (measured in PSNR).
		}
		\label{compare}
	\end{figure*}

The gain in PSNR is \SI{1.0}{\decibel}, which makes up for about \SI{46}{\percent} of the maximal gain of \SI{2.2}{\decibel}.
Reconstructions obtained from the optimal uncorrected ($n=29$ iterations of ML-EM are used on gate zero) and the proposed method with the optimal number of iterations of ML-EM and \MMLEM are presented in \autoref{compare}.
The proposed method seems to give smoother results.
The smaller discs towards the middle of the image are also better seen.

We also emphasise that these results (improvement in PSNR and optimal number of iterations) are extremely robust with respect to the randomness involved in the experiments, namely the vector fields drawn randomly as well as the Poisson noise.

\subsection{Implementation Details}

All computations are run in Python and use Operator Discretization Library (\texttt{odl}) for manipulating operators~\cite{Adler2017odl}, \texttt{neuron} for warping utilities~\cite{Dalca2018anat}, which itself uses \texttt{tensorflow}~\cite{Abadi2016}.
The training was performed with \texttt{voxelmorph}~\cite{Dalca2018}.

%


\section{Perspectives}
This paper presents a new method for joint motion estimation and image reconstruction in PET.
Its main advantage is its cost, similar to that of the usual ML-EM algorithm, making it scalable to clinical 4D data.

Our framework also allows for further modelling such as attenuation correction. 
In a future work, we consider testing this method with clinical data.
This would require training the network on appropriate datasets.
We also plan to generalise the approach to other group actions, such as the mass-preserving one, which is more physically relevant.

%
%

\subsection*{Acknowledgements} 
We acknowledge support from the Swedish Foundation of Strategic Research grant AM13-004.

\bibliographystyle{splncs04}

\bibliography{biblio,biblio2}

\end{document}